\documentclass[12pt,a4paper,final]{iopart}
\usepackage{iopams}  
\usepackage[dvips]{graphicx}
\usepackage[breaklinks=true,colorlinks=true,linkcolor=blue,urlcolor=blue,citecolor=blue]{hyperref}

\begin{document}

\title[Pattern Transitions in Unstable Viscous Convective Medium ]{Pattern Transitions in Unstable Viscous Convective Medium }

\author{I.V. Gushchin, A.V. Kirichok, V.M. Kuklin} 

\address{Kharkov National University, Institute for High Technologies,\\ 4 Svobody Sq., Kharkov 61077, Ukraine}

\ead{kuklinvm1@rambler.ru}

\begin{abstract}
Convection in a thin layer of liquid (gas) with temperature dependent viscosity between poorly heat conducting boundaries is studied within framework of the Proctor-Sivashinsky model. This model is examined in order to study both the flow pattern formation and the second-order structural phase transitions as between patterns with translational invariance as well as between structures with broken translational invariance but keeping a long-range order. The spatial spectrum of arising patterns and estimation of their visual defectiveness are analyzed. The relation between the density of pattern defects and spectral characteristics of the pattern is found. We also discuss the noise effects on the formation of pattern defects. The influence of temperature dependence of viscosity on the process of pattern formation and structure transformations is also discussed. It is shown that the temperature dependence of viscosity inhibits structural transition from regular rolls to square cells.
\end{abstract}

\vspace{2pc}


\section{Introduction}

The mechanisms of pattern transformations and second-order structural-phase transitions between different patterns that result in changes of their symmetry and partly of their characteristic scales have always been of great interest to researchers and developers of technologies.

Considering the various processes in continuous media, we need to take into account the dynamics of perturbations with not only different spatial and temporal scales but also with different spatial orientation \cite{Busse 1980,Chandrasekhar 1970,Getling 1998,Cross 1993,Karpman 1973,Engelbrecht 1985,Schwartz 1982,Davydov 1984,Yanovsky 2007,Zaslavsky 1991,Kuklin 2004}. The last one is responsible in the common geometric sense for the spatial structure symmetry, which possess not only short-range but also long-range order \cite{Nikolis 1977,Haken 1978}.

The nonlinearity of the medium manifests itself in certain mechanisms of interaction between these perturbations. Different approaches to the description of such interaction in nonequilibrium media are presented in \cite{Tsytovich 1970,Zakharov 2012,Petviashvili 1989,Klimontovich 1990,Lvov 1994,Kuklin 1989,Buts 2006,Moiseev 1988}. The processes involving a large number of perturbations of all scales and orientation are often called multi-wave or multimode when the matter concerns wave media or periodic systems.

Currently, the issue of the greatest interest is the elucidation of the nature of spatial structures formation, the search for physically transparent mechanisms of these processes, and then formulation of adequate (possessing a clear physical background) mathematical models for description of these phenomena. 

Considering the behaviour of multimode or multiwave spatial structures formation processes, we can see the appearance of their specific features, such as a change in dynamics of the instability (i.e. delay or even suppression \cite{Horsthemke 1984}) during the formation of unstable nonlinear structures. The number of degrees of freedom plays the significant role as well as the number of spectrum modes that allows to use a small parameter $\eta \propto 1/N$ inversely proportional to the number of spectrum modes $N$). The existence of a dense spectrum of perturbations can form long-lived metastable nonlinear states and can delay the development of transient processes and structural phase transitions between these states \cite{Kirichok 1999,Belkin 2010}.

The Proctor-Sivashinsky model is found to be very attractive \cite{Chapman 1980,Gertsberg 1981} for studying the processes of pattern formation in systems which possess a preferred characteristic spatial scale of interaction between the elements of future structure. This model was developed for description of convection in a thin layer of liquid between poorly conducting horizontal boundaries. Authors of \cite{Malomed 1989} have found the stationary solutions with a small number of the spatial modes, one of which (convective cells) was steady and the second one turned out to be unstable (convective rolls). A particular future of the model is that it forces a preferred spatial scale of interaction, leaving the system a chance of selecting the symmetry during evolution. It was found, that the type of symmetry and hence the characteristics of the structure are determined by the minima of the potential of interaction between modes lying on a circle in $\vec k$-space. A short time later, Pismen \cite{Pismen 1986} upgraded the Proctor-Sivashinsky model by including the inertial effects and taking into account the poloidal vortices inside the layer. This model, as was shown by further researches, enables the correct description of energy transfer from toroidal Proctor-Sivashinsky vortices (which forms the periodic structure) to large-scale poloidal vortex motion \cite{Kirichok 1993,Kirichok 1997}. This phenomenon of the "hydrodynamic dynamo" is may be responsible for the formation of large-scale vortices in convective layers, in particular in the atmosphere of planets. However, even within the Proctor-Sivashinsky model not all processes and the phenomena were studied.

The detailed analysis of instability leading to the formation of a metastable structure (convective rolls) will be presented below. Earlier, it was found \cite{Kirichok 1999} that at first stage of the instability evolution the metastable long-lived state (the curved quasi-one-dimensional convective rolls) arises. And later, after a lapse of time (which is considerably greater than the reverse linear increment of the process), the system transforms to the steady state (square convective cells). The detailed treatment of the Proctor--Sivashinsky model \cite{Belkin 2010} presented below shows that this structural transition demonstrates all the characteristics of the second order phase transition (the continuity of the sum of squared mode amplitudes over the spectrum $I=\sum _{j}a_{j}^{2}  \equiv \sum _{k_{j} }|a_{k_{j} }^{} |^{2}  $or that the same, the continuity of density of this value and discontinuity of its time derivative $\partial I/\partial t$ ). The crucial issue discussed in this paper is the determination of the degree of defectiveness (pattern imperfection) of originating regular structures and also the searching for a correlation between integral spectral characteristics and a fraction of defective cells in the structure. The defectiveness of the structure appears, in particular, in the intermediate transient regimes and is caused by stimulated (due to non-equilibrium) interference of growing modes \cite{Kuklin 2006}. In the case of the external influence, the noise is able to support the set of weak spatial modes which were suppressed before and which interfere with dominating modes is also capable to provide the interference pattern corresponding to the imperfect spatial pattern. The understanding of processes which lead to violations of spatial periodicity of the structure, would allow estimating the level of pattern imperfection by their spatial spectrum that can be quite possible measured experimentally. Especially, it should be clarified the influence of external noise on stability of states and structural-phase transitions.

 The existence of preferred scale (the distance between the regular spatial perturbations) and the possibility to select the type of symmetry (the regular spatial configuration) motivate the interest to this physical model, particularly for description of processes in solid state physics, where the characteristic distance between elements of spatial structures (atoms, molecules) in their condensed state is almost invariable. 

It is also shown below that the intermediate states with broken short-range order, but keeping a long-range order can appear as a result of structural-phase transitions (second order phase transitions) and demonstrate the same formation dynamics, as the regular spatial structures.

The objective of this work is investigation of the mechanisms of pattern formation and mode competition in convective medium. The nature and evolution of structural phase transitions between patterns of different topology are considered in detail. Besides the regular periodic structures, we also analyse the imperfect patterns i.e. the structures with implemented spatial defects.

\section{Model description}

When the Rayleigh number $Ra$ exceeds the critical value $Ra_{c}$ corresponding to the onset of convective flow, i.e. $Ra=Ra_{c} (1+\varepsilon )$ ($\varepsilon\ll 1$), the three-dimensional convection begins in a thin layer of liquid between poorly conducting horizontal plates heated from below (see, for example \cite{Chandrasekhar 1970}), which can be described by the Proctor-Sivashinsky equation \cite{Chapman 1980,Gertsberg 1981}. This equation determines the evolution of  the  "order  parameter"  $\Phi$,  which  has  a  meaning of  a  temperature  deviation  averaged  in  the  vertical direction: 

\begin{equation} \label{eqn1}
\dot{\Phi }=\varepsilon ^{2} \Phi +\gamma \nabla (\Phi \nabla \Phi )-(1-\nabla ^{2} )^{2} \Phi +\frac{1}{3} \nabla \left(\nabla \Phi \left|\Phi \right|^{2} \right)+\varepsilon ^{2} f 
\end{equation}

\noindent Here,  the  dot  indicates  differentiation with  respect  to time,  and $\nabla$ is  the  two-dimensional vector  differential  operator in the horizontal plane $(X, Y)$, $f$ is the random function describing the external noise.
The term $\gamma \nabla (\Phi \nabla \Phi )$ describes the temperature dependence of viscosity.
Also, we use the following definitions: $X\approx 4x\sqrt{\varepsilon }$, $Y\approx 4y$, $F\approx 4\Phi $ is the dimensionless temperature in terms of the difference between fixed temperature of the bottom plane $T_{d} $ and the upper surface $T_{u} $ in the absence of convection; $\psi$ is the dimensionless velocity in units of the thermal diffusivity $\chi$ (is equal to the coefficient of thermal conductivity $\lambda $, divided by the density of $\rho$ and thermal conductivity $C_{p} $) of fluid; coordinates $x$, $y$ in terms of the layer thickness $d$, time interval $d^{2} /\chi $, the Rayleigh number $Ra=\sigma g(T_{d} -T_{u})d^{3}/ \nu \chi$ is the number that determines the behaviour of the fluid under the action of a temperature gradient (convection currents arise when this parameter exceeds the threshold value ); $\nu$ is the kinematic viscosity (dynamic viscosity of the kinematic, multiplied by the density), $\sigma$ here is the coefficient of thermal expansion of the liquid.

\begin{table}[t]
\caption{\label{tab1}The correspondence of used variables and their real physical values \cite{Gertsberg 1981}}
\begin{indented}
\item[]\begin{tabular}{@{}ll}
\br
Physical quantity & Representation of explicit view \\
\mr
Temperature $T\left(x\sqrt{\varepsilon } ,y\right)$ & $T_{d} +\left(T_{d} -T_{u} \right)\left(-y+F(x\sqrt{\varepsilon } ,y\right)$ \\ 
Horizontal velocity $\psi _{y} $ & $60\sqrt{\varepsilon } \cdot F_{x\sqrt{\varepsilon } } \cdot \left(2y^{3} -3y^{2} +y\right)$ \\ 
Vertical velocity $-\psi _{x} $ & $-30\varepsilon \cdot \left(F_{x\sqrt{\varepsilon } } \right)_{x\sqrt{\varepsilon } } \cdot \left(y^{4} -2y^{3} +y^{2} \right)$ \\ 
\br 
\end{tabular}
\end{indented}
\end{table}

We seek a solution of (\ref{eqn1}) in the form 

\begin{equation} \label{eqn2} 
\Phi =\varepsilon \sum _{j}a_{j} \exp (i\vec{k}_{j}\cdot \vec{r}),
\end{equation}

\noindent with $|\vec{k}_{j} |=1$. After renormalization of time units $\propto \varepsilon ^{2} $, we obtain the evolution equation for slow amplitudes $a_j$ [30]:

\begin{equation} \label{eqn3}
\dot{a}_{j} =a_{j} -2\gamma a_{j+j_{0} } a_{j+2j_{0} } -\sum _{m=1}^{N}V_{mj} |a_{m} |^{2} a_{j}  +f
\end{equation}

\noindent where interaction coefficients are determined as follows

\begin{equation} \label{eqn4} 
V_{jj}=1,
\end{equation}
\begin{equation} \label{eqn5} 
V_{ij} =\frac{2}{3}(1+2(\vec{k}_{i} \vec{k}_{j})^{2})=\frac{2}{3}\left(1+2\cos ^{2} \vartheta \right).  
\end{equation}

\noindent Here $\vartheta $ is the angle between vectors $\vec{k}_{i} $ and $\vec{k}_{j} $.  Let  $\vartheta _{j_{0} } =2\pi /3$, $\vartheta _{j+j_{0} } =\vartheta _{j} +2\pi /3$  and  $\vartheta _{j+2j_{0} } =\vartheta _{j} +4\pi /3$.       

The instability interval in $k$-space represents a ring with unit average radius and the width of order of relative above-threshold parameter $\varepsilon$, i.e. much less than unity. During the development of the instability, the effective growth rate of modes that lies outside of the very small neighbourhood near the unit circle will decrease due to the growth of the nonlinear terms and can change sign which will lead to a narrowing of the spectrum to the unit circle in the $k$-space. Since the purpose of further research will be the study of stability of spatial structures with characteristic size of order $2\pi /k\propto 2\pi $ and the important characteristic for visualization of simulation results will be evidence of these structures, so we restrict ourselves by considering some idealized model of the phenomenon, assuming that the oscillation spectrum is already located on the unit circle in the $k$-space.

\section{Simulation results}

\textbf{\textit{Convection with temperature independent viscosity.}} It was shown in \cite{Kirichok 1999, Belkin 2010} that in the absence of temperature dependence of viscosity and when the number of modes is sufficiently large, the system delayed the development while remaining in a dynamic equilibrium. Development of perturbations in the system, as shown by the numerical analysis, will be as follows \cite{Kirichok 1999}. 

Starting from initial fluctuations, the modes over a wide range of $\vartheta$ begin grow. The value of the quadratic form of the spectrum $I=\sum _{j} a_{j}^{2}$ after the first peak of the derivative can be estimated as a value close to 0.75. This is so-called "amorphous" state --  the first metastable structure.

For further development - "crystallization", one of the modes must get a portion of the energy which excesses some threshold value. That is, in these case, it is necessary a certain level of noise (fluctuations). This can be achieved either at finite noise level$f\ne 0$ or by decreasing the accuracy of calculations that is the same as noted in \cite{Belkin 2010}. The similar cases, when the noise can trigger or accelerate instability, are reviewed in the book \cite{Horsthemke 1984}.

If one of the modes gets the proper amount of energy, then the process of formation of a simplest convective structure -- rolls begins. Note that in the nature, the thin clouds also can form the roll structure. The value of $I$ in this case tends to unity ($I\to 1$). However, this state is not stable and then we can see the next structural transition: convective rolls are modulated along the axis of fluid rotation, and the typical size of this modulation phases down. In this transition state, the system stays for a sufficiently long time (which slightly increases within some limits with increase in the number of modes), and the value $I\approx 1.07$ remains constant during this time. After a rather long time, ten times more than the inverse linear growth rate of the initial instability only the one mode ``survives'' from newly formed ``side'' spectrum, which amplitude is comparable with the amplitude of the primary leading mode. In the end, the stable convective structure -- square cells is generated, and the quadratic form $I$ reaches the value of $I=1.2$.

Further researches of this process have found the following dynamics of quadratic form $I=\sum _{j}a_{j}^{2}$ with time (see Fig.2).  Near the second peak of the derivative $\partial I/\partial t$, the metastable structure -- a system of convective rolls with $I\approx 1$ is formed, and remains unchanged up to the moment when the second splash of $\partial I/\partial t$ have appeared that indicates the emergence of a secondary metastable structure -- transversally modulated imperfect rolls with a new value of $I\approx 1.07$. 

\begin{figure}
\center{\includegraphics[width=7cm]{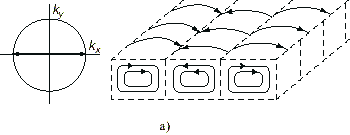}
\includegraphics[width=7cm]{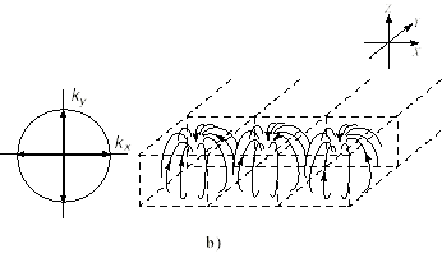}}
\caption{Convective structures: rolls (a) and square cells (b).}
\label{fig1}
\end{figure}

After the second splash of the time-derivative, a stable structure of squared convective cells is started to build up. Such behaviour proves the existence of structural-phase transitions in the system.

\begin{figure}
\center{\includegraphics [width=10cm] {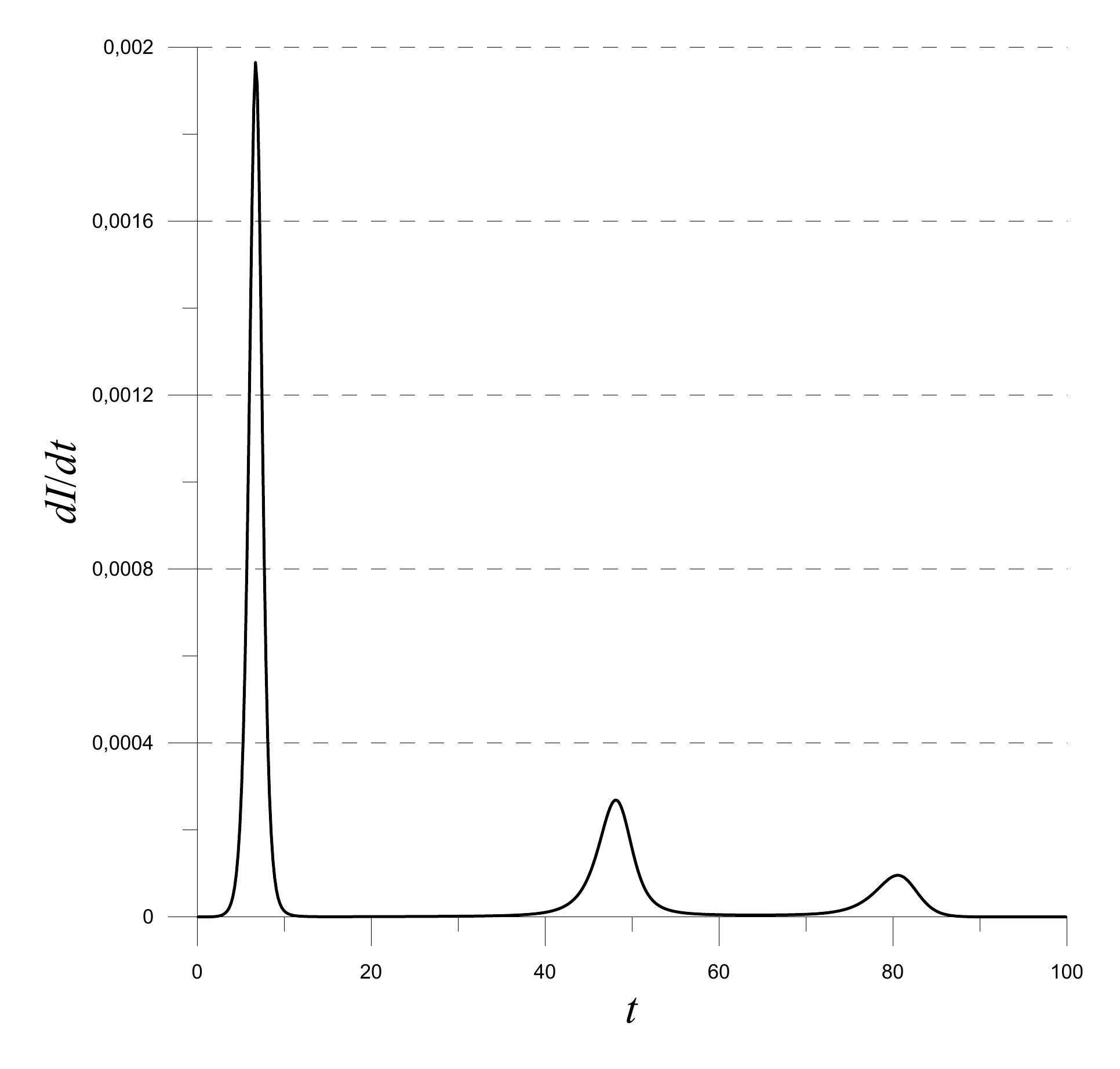}}
\caption{The evolution of the derivative $dI/dt$ (in relative measuring units) of the integral quadratic form $I=\sum_j a_j^2$.}
\label{fig2}
\end{figure}

Generally speaking, the characteristic times of relaxation processes during evolution of the system to more equilibrium state are determined as usual by the difference of the state function values before the transition and after it. The greater this difference, the faster the transition from one state to another.

It is important to keep in mind that the sequence of state transitions is determined by the characteristic times of instabilities (which play the role of relaxation processes) that provide a cascade evolution of the system to the most equilibrium state. Initially, the most fast relaxation processes take place, that associated with large difference of the state function values corresponding to different equilibrium states. 

Let us verify that in this case all the phenomena occur in the same order and within the framework of the foregoing scenario. The numerical analysis of the model allows to confirm these considerations.

It can be seen that the times of state formation $\tau _{n} $ are inversely proportional to the difference between the values $I=\sum _{i}A^{2} _{_{i} }  $ after n-th structural phase transition $I_{n}^{(+)} =(\sum _{i}A^{2} _{_{i} }  )_{n}^{(+)} $   and before it $I_{n}^{(-)} =(\sum _{i}A^{2} _{_{i} }  )_{n}^{(-)} $, i.e.

\begin{equation} \label{eqn6} 
\tau _{n} \propto \left\{ \left(\sum _{i}A^{2} _{_{i} }  \right)_{n}^{(+)} -\left(\sum _{i}A^{2} _{_{i} }  \right)_{n}^{(-)} \right\} ^{-1} =\Delta I_{n}^{-1}  
\end{equation} 
   It follows from this that
\begin{equation} \label{eqn7} 
\tau _{3} /\tau _{2} \approx \Delta I_{2} /\Delta I_{3} ,                                                                                                               
\end{equation} 

Thus, we have shown by numerical simulation of the Proctor-Sivashinsky model that the state with certain topology can be described by the state function, which is the sum of squared mode amplitudes $I=\sum _{i}A^{2} _{_{i} }  $. The transitions between these states are characterized by splashes in time-derivative of this function and different meta-stable structures, corresponding to different values of the state function have different visually distinguishable topologies.

Let us consider in more detail the formation of square convective cells. Denote the amplitudes of the modes forming a spatial structure of square convective cells as $a_{1} $ and $a_{2} $. Consider the dynamics of "spectrum defectiveness (imperfection)" parameter of the structure $D=\sum _{j\ne 1,2}a_{j}^{2}  /\sum _{j}a_{j}^{2}  $. It is defined as the ratio of the sum of squared mode amplitudes which does not fit the system of square cells to the total sum of modes squares. In addition, let introduce so-called ``visual defectiveness  (imperfection)'' parameter $d=N_{def} /N$, where $N_{def} $ is the number of defective spatial cells (the area of the structure occupied by irregular cells) and $N$is the number of cells in a perfect regular structure (the total area of the structure). The process of structure rearrangement is observed in the interval between the second and third splashes of the derivative quadratic form (\Fref{fig2}).

\begin{figure}
\center{\includegraphics[width=10cm] {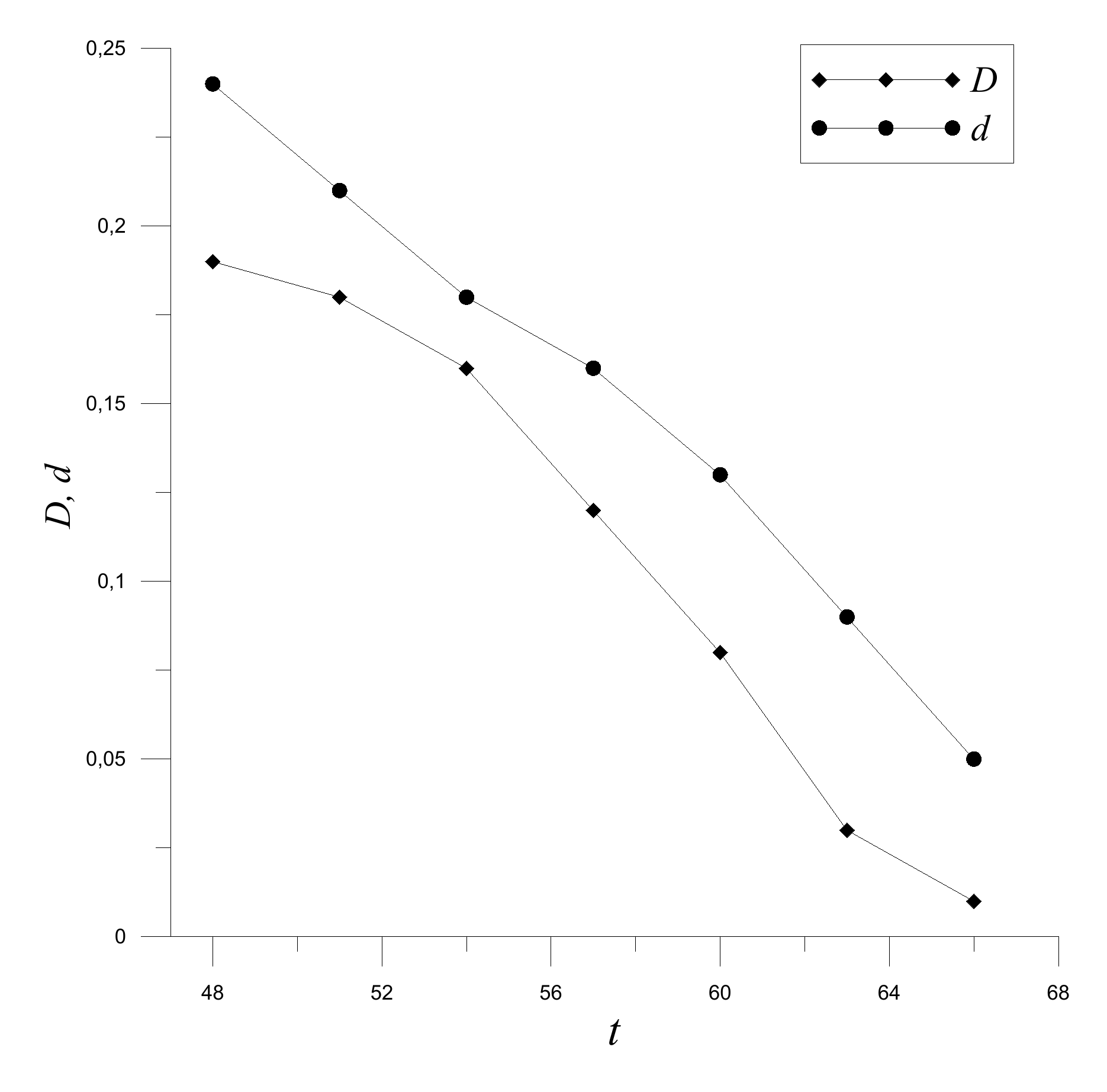}}
\caption{Comparative analysis of the spectral defectiveness $D$ and visual defectiveness $d$ of the structure. The number of modes is 50.}
\label{fig3}
\end{figure}

The criteria by which the cell was considered as regular and the method of calculation the number of these cells are following. The picture for the field is converted to 8-bit image, i.e. the maximum number of colours is reduced to 256. Thus, the formed structure becomes more evident and observable. Increasing this image, one can quite clearly distinguish which of the structural units is the proper cell, and which is not. The proper cell has the correct geometry with uniformly dark center and four lighter hills surrounding the center and of comparable size. Despite on qualitative character of quantity description characterizing the spectral and visual defectiveness of structure we can note a similarity in its behaviour (see \Fref{fig3}) near the completion of the structural transition.

In the case of a sufficiently high level of noise, both additive ($f\ne 0$) and multiplicative (a random component proportional to $\varepsilon^{2} $ in the first term in r.h.s of Eq. \eref{eqn1}), the level of modes amplitudes may be rather large from the start of the process. The initial conditions may also provide the starting system state can be considered as highly irregularity "amorphous", i.e. the perturbation amplitudes are large enough and randomly different from each other. This state can be maintained in the future by random noise. It is important to find out in what noise levels it is possible the "amorphous" state, characterized by a large number of spatial modes can exist for a long time. 

Apparently, the very intensive noise is able to keep the system from the formation of convective structures, however, preliminary estimates suggest that the noise of lesser intensity cannot prevent the successive transition to metastable (rolls) and stable (square cells) states. When the noise intensity falls down, the transition from a metastable to stable state can slow down and the system stays ("freezes") for a long time in the metastable state.

\begin{figure}
\center{\includegraphics [width=10cm]{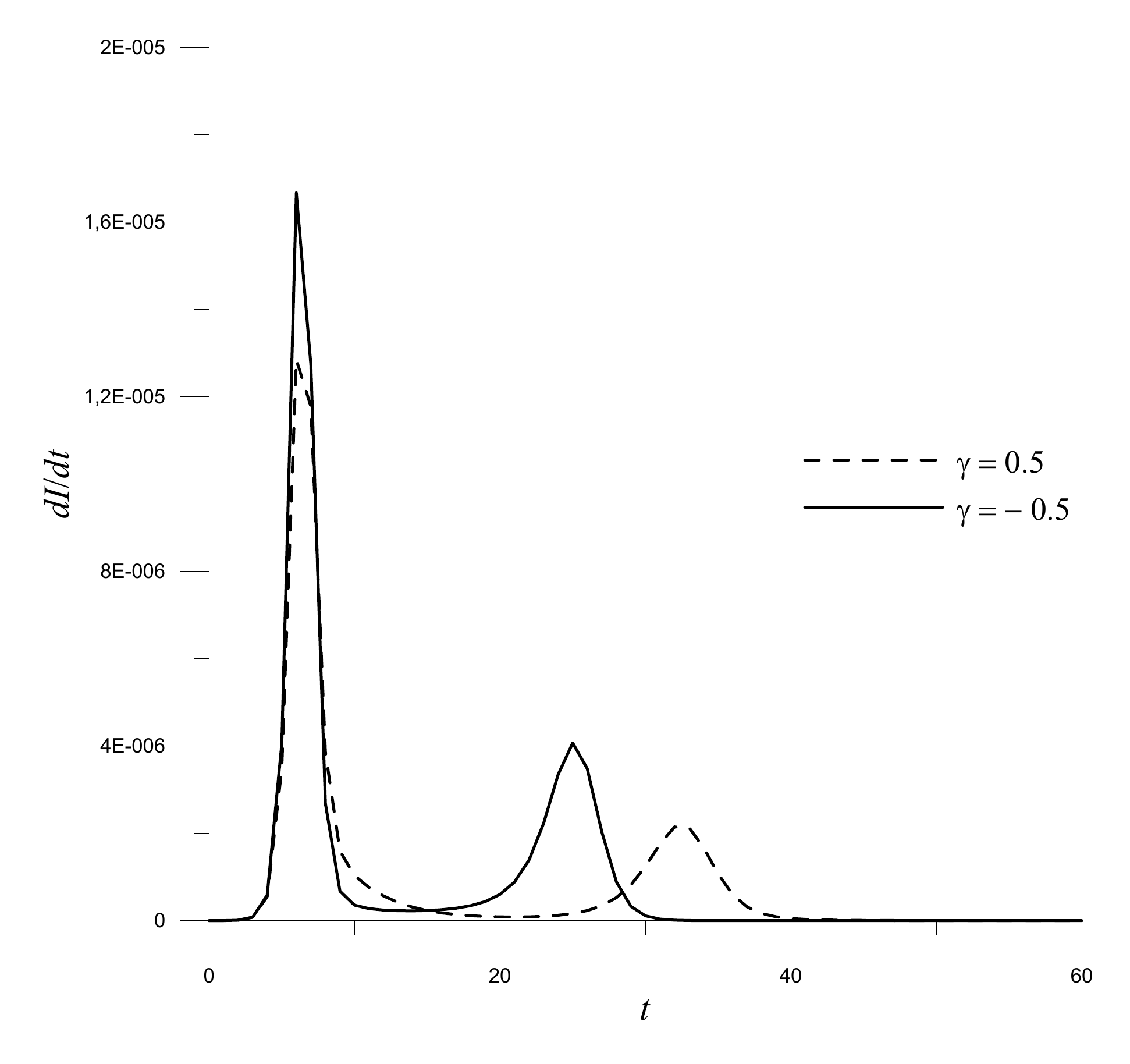}}
\caption{The evolution of the derivative $\partial I/\partial t$ for $\gamma = \pm 0.5$.}
\label{fig4}
\end{figure}

\begin{figure}
\center{\includegraphics [width=10cm]{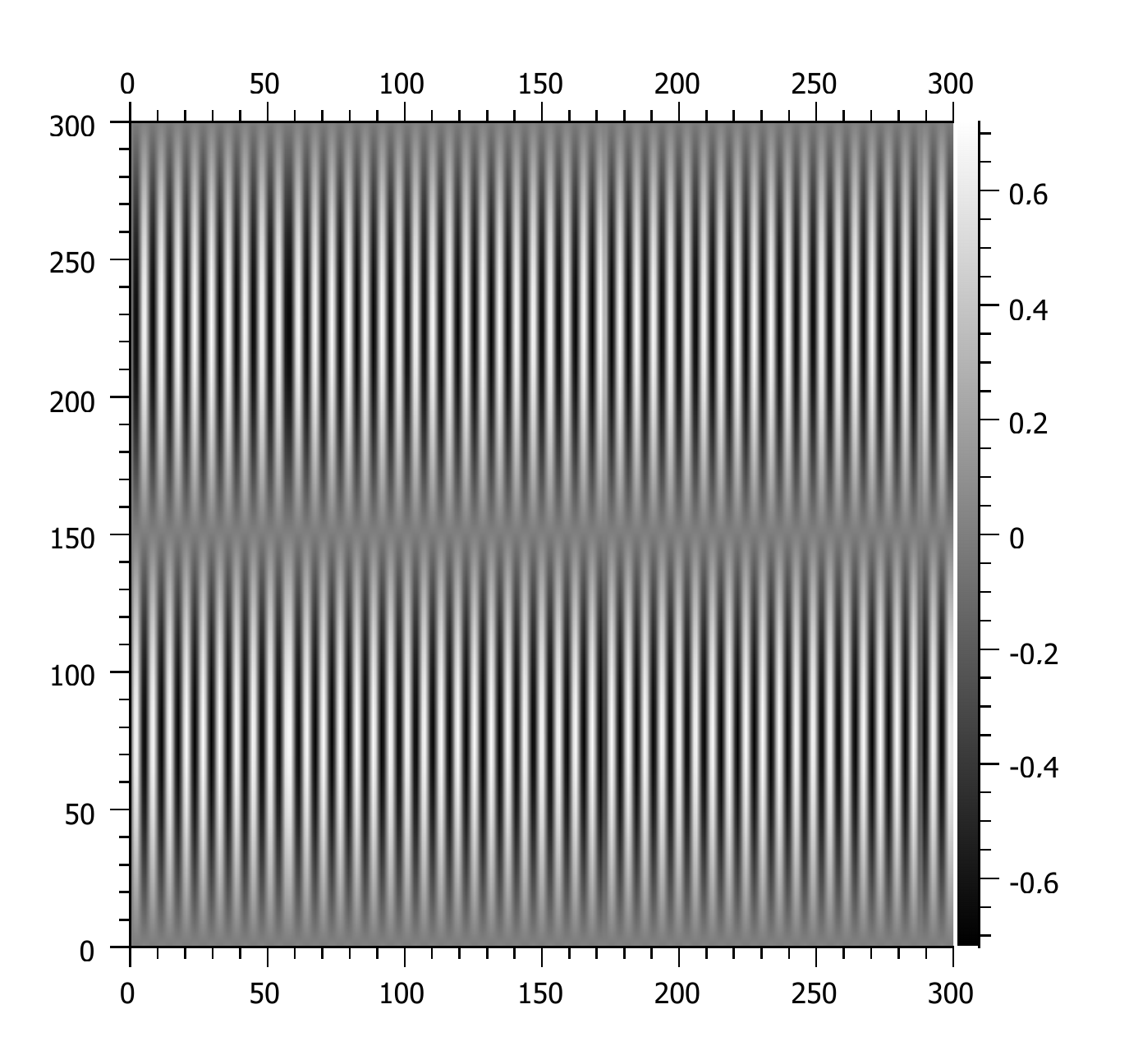}}
\caption{Temperature field corresponding to the roll structure, $\gamma =\pm 0.5$.}
\label{fig5}
\end{figure}

\textbf{\textit{Convection with temperature dependent viscosity. }}The term $\gamma \nabla (\Phi \nabla \Phi )$ appears in Eq. \eref{eqn3} when we take into account the temperature dependence of viscosity. For $\gamma >0$ the gas (this case corresponds to the gas convection) flows up to the center of the cell, for $\gamma <0$ (which corresponds to the movement of the liquid) the liquid flows outward and down from the center of the cell (see for example \cite{Gushchin 2013}).  At $|\gamma |<<1$, the effect of this term on the dynamics of the process is negligible. The convection develops in accordance with above-described scenario. However, when the parameter $\gamma $ approaches the unit value, one further mechanism of energy transfer between each triplet of interacting modes appears which destroys the previous mechanism of mode interaction arising due to vector cubic nonlinearity.  The consequences of this destruction are almost identical for $\gamma $ of different signs.  First of all, the rapid growth of the modes spectrum at the linear stage of instability forms a metastable structure with rather intricate topology, depending on the initial conditions. However, after a short time there is a second structural transition (\Fref{fig4}) as a result of which  stable and well-defined elongated rolls are formed,  which structure is shown in \Fref{fig1}. The spatial distribution of temperature field of the structure is demonstrated in \Fref{fig5}. 

\begin{figure}
\center{\includegraphics [width=10cm]{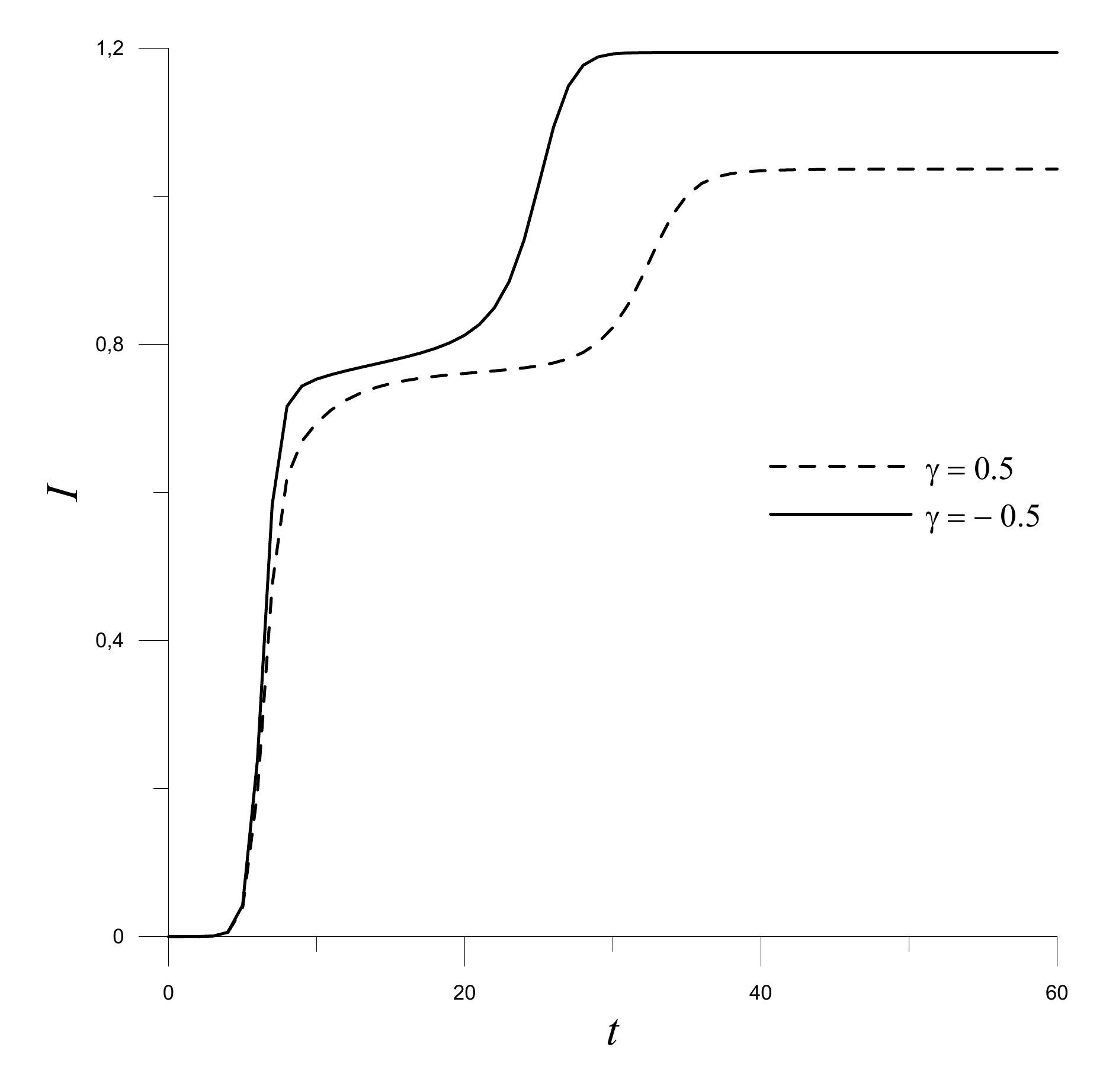}}
\caption{ Dynamics of the quadratic form $I=\sum _{j}a_{j}^{2}  $, which characterizes the state of the system at $\gamma =\pm 0.5$.}
\label{fig6}
\end{figure}

\begin{figure}
\center{\includegraphics[width=10cm] {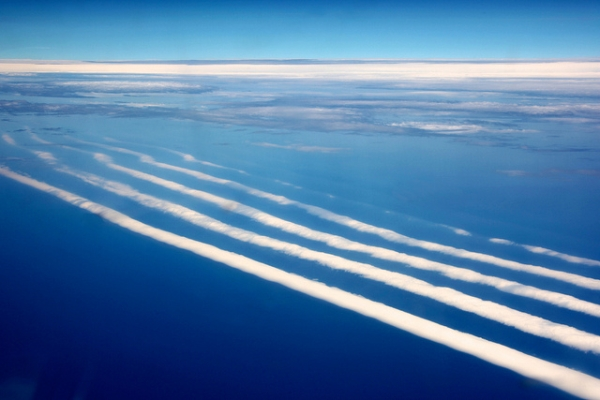}}
\caption{ Formation of convective rolls, extending for hundreds of kilometres in the north of Australia at the beginning of rainy season.}
\label{fig7}
\end{figure}

\noindent \Fref{fig6} demonstrates the specific features of structural transitions, where one can see the regularity of the function $I=\sum _{j}a_{j}^{2}  $, which characterizes the state of the system.  Thus, an appreciable temperature dependence of viscosity can lead to formation of stable convective rolls. Such convective rolls can be observed in the thin cloud cover (\Fref{fig7}). 

\noindent 

\section{Conclusion}

The special feature of the Proctor-Sivashinsky model with temperature independent viscosity is the existence of three possible metastable states. The times of structural transitions between these metastable states are much less than the times of their existence.

Each metastable state has a definite topology and can be characterized by definite steady value of the state function $I=\sum _{i}A^{2} _{_{i} }$. The metastable states are destroyed with time for the instabilities, the growth rate of which can be evaluated from the amplitude of splashes of  time-derivative of the state function. 

It is shown that the characteristic times of the instabilities, which destroy the previous and form a new state are inversely proportional to the difference between the values of the state function before and after the structural phase transition. In addition, we show that the faster relaxation processes, i.e. structural phase transitions take priority over more slow ones.

The characteristic size of the convective structures in the regime of advanced instability is of order $2\pi /k\propto 2\pi$ and the length of the wave vectors is of order unity (in conventional dimensionless units). The potential of interaction between spatial modes $V_{ij} {\rm =}{(2\mathord{\left/ {\vphantom {(2 3)\left(1+2\cos ^{2} \vartheta _{ij} \right)}} \right. \kern-\nulldelimiterspace} 3)\left(1+2\cos ^{2} \vartheta _{ij} \right)} $ has a deep minimum for angles $\vartheta _{ij} =\vartheta _{i} -\vartheta _{j} =\pm {\pi \mathord{\left/ {\vphantom {\pi  2}} \right. \kern-\nulldelimiterspace} 2} $ between vectors $\vec{k}_{i} $ and $\vec{k}_{j} $. Namely, these minima are the reason of the instability of convective rolls \cite{Kirichok 1999,Belkin 2010}, because the existence of a minimum $V_{ij} $ for modes with relatively low amplitudes allows them to continue growth, while suppressing the perturbations occurred before.

When approaching to the stable state, the spatial structure gets rid of many defects. There is a correlation between the relative fraction of visually (geometrically) observed structural defects and the defectiveness parameter, defined as the ratio of the squares of the spectrum mode amplitudes, which does not fit the system of square cells, to the total sum of squared mode amplitudes.

The temperature dependence of viscosity, included in the Proctor-Sivashinsky model, results in the suppression of structural phase transition, which previously led to the formation of square cell pattern. As in the absence of temperature dependent viscosity, the long-lived metastable states with a topology that is defined by the boundaries of the system and the initial conditions are observed. Some differences between the gas and liquid media consist only in small difference in the amplitude of the final structure of the convective rolls, without changing the nature of structural phase transitions. 

 The authors thank prof. Parkhomenko A.A., providing us a picture of clouds over Australia. 

\bibliographystyle{plain}

\end{document}